\documentclass[iopat,jpcm,twocolumn,amsmath,amssymb,superscriptaddress,showpacs]{revtex4-1}

\usepackage[utf8]{inputenc}
\usepackage{amsfonts}
\usepackage{amsmath}
\usepackage{amssymb}
\usepackage{amsthm}
\usepackage{fontenc}
\usepackage{graphicx}
\usepackage{xcolor}
\usepackage{textcomp}
\usepackage{epstopdf}
\usepackage{braket}
\usepackage{mathtools}
\usepackage{amsmath}
\usepackage{dcolumn}

\begin{document}
\newcommand{\abs}[1]{\lvert#1\rvert}
\title{Fano resonances in hexagonal zigzag graphene rings under external magnetic flux} 


\author{D. Faria}
\email{daiara@if.uff.br}
\affiliation{Universidade Federal Fluminense,  Av. Litor\^{a}nea sn, 24210-340 Niter\'{o}i, RJ, Brasil}
\author{R. Carrillo-Bastos}
\affiliation{Ohio
University, Athens, Ohio 45701-2979, USA}
\affiliation{Centro de Investigaci\'{o}n Cient\'{i}fica y de
Educaci\'{o}n Superior de Ensenada, Apado. Postal 360, 22800
Ensenada, Baja California, M\'{e}xico} \affiliation{Universidad
Nacional Aut\'{o}noma de M\'{e}xico, Apdo. Postal 14, 22800
Ensenada, Baja California, M\'{e}xico } 
\author{N. Sandler}
\affiliation{Ohio University, Athens, Ohio 45701-2979, USA}
\author{A. Latg\'{e}}
\affiliation{Universidade Federal Fluminense,  Av. Litor\^{a}nea sn, 24210-340 Niter\'{o}i, RJ, Brasil}

\date{\today}

\begin{abstract}
We study transport properties of hexagonal zigzag graphene quantum rings connected to semi-infinite nanoribbons. Open two-fold symmetric structures support localized states that can be traced back to those existing in the isolated six-fold symmetric rings. Using a tight-binding Hamiltonian within the Green's function formalism, we show that an external magnetic field promotes these localized states to Fano resonances with robust signatures in transport. Local density of states and current distributions of the resonant states are calculated as a function of the magnetic flux intensity. For structures on corrugated substrates we analyze the effect of strain by including an out-of-plane centro-symmetric deformation in the model. We show that small strains shift the resonance positions  without further modifications, while high strains introduce new ones. 
\end{abstract}


\maketitle

\section{Introduction}
The unique electronic properties of graphene\cite{Castro} have guided research combining confinement and interference effects on annular systems\cite{Trauzettel}. Closed ring geometries for example, have been extensively studied, rendering predictions of energy spectra for different choices of boundary conditions \cite{Morpurgo,Peter, Xu, Peeters,Brey} and including the role of strain\cite{Daiara}. When a magnetic flux is applied in these Aharonov-Bohm geometries, persistent currents with lifted-valley degeneracy appear\cite{Morpurgo,Wurm}. Models with different ring geometries reveal peculiar transport properties: rectangular shapes exhibit resonant tunneling phenomena\cite{Peeters1}, while hexagonal ones \cite{Beenakker} allow for current blocking mechanisms with the leads acting as valley filters. Samples with some of these characteristics have already been synthezised\cite{Russo, Ihn1, Ihn2, Haug, Rahman, Haug2}, with some fabrication techniques able to produce rings with perfect hexagonal symmetry by exploiting appropriate lattice orientations\cite{Hu,Baringhaus}. Transport measurements in these open rings show peculiar conductance oscillations in the presence of magnetic fields\cite{Haug2}. These new devices present the opportunity to test theoretical predictions and reveal new transport phenomena  that may be used to develop new technological applications. 

Ring geometries are particularly useful to investigate interference effects that may appear with precise fingerprints, such as those produced by Fano resonances\cite{Miro, Boris,FanoAharonov}. Fano physics is a rich phenomena produced by the coexistence of resonant (localized) and nonresonant paths for scattering waves\cite{Fano}. The transmission function exhibits an asymmetric line shape which depends on the coupling between resonant and extended states. In particular, Fano resonances have been reported in semiconducting ring shaped structures under the presence of an external magnetic field \cite{Szafran}. The resonances arise as consequence of the broken time-reversal symmetry of localized states and their interference with extended states through the semiconducting ring. Fano resonances were also observed in similar semiconducting rings as a consequence of broken parity of localized states due to spin-orbit interaction\cite{Nowak}. In semiconductors, emergent bound states are produced by disorder that could create states in the gap (no observed Fano effect). It is important to mention that clean semiconductor samples will not present Fano resonances unless some symmetry is broken in the system. In graphene systems, various interesting proposals for observing Fano resonances have been advanced\cite{Gonzales,Adame,Adame1,Rosales}, with some ring geometries exploiting a broken upper-lower arm symmetry to produce the effect\cite{Adame}.  Transmission calculations in different hexagonal configurations were explored and the conductance bands of zigzag rings were found to be considerably narrower than in the case of armchair edges, being less robust under perturbations\cite{Adame}.

In this paper, we carry out a study of transport properties of a graphene quantum ring connected to semi-infinite zigzag nanoribbons that exhibits signatures of Fano interference under external magnetic flux. The isolated structure possesses a hexagonal (six-fold) symmetry with zigzag inner and outer edges. Connecting the ring to leads renders a  two-fold symmetric setup  that  retains localized states as shown in density of states (DOS) calculations (see below). We show that an external magnetic field that fully pierces the ring region produces signatures of Fano resonances in the DOS and strongly modifies the ring conductance in a wide range of energies. These resonances are a result of the pre-existing localized states that couple to the continuum by the external flux.
The addition of an out-of plane centro-symmetric deformation in the ring region, described by an effective pseudo-magnetic field\cite{Vozmediano}, modifies the transmission properties of the structure. The resulting Fano resonance energies are shifted from their original values  for small strains while new resonances appear at energies precisely determined by the curvature of the deformation. 

\begin{figure}[h!]
\centering
\includegraphics[scale=0.51]{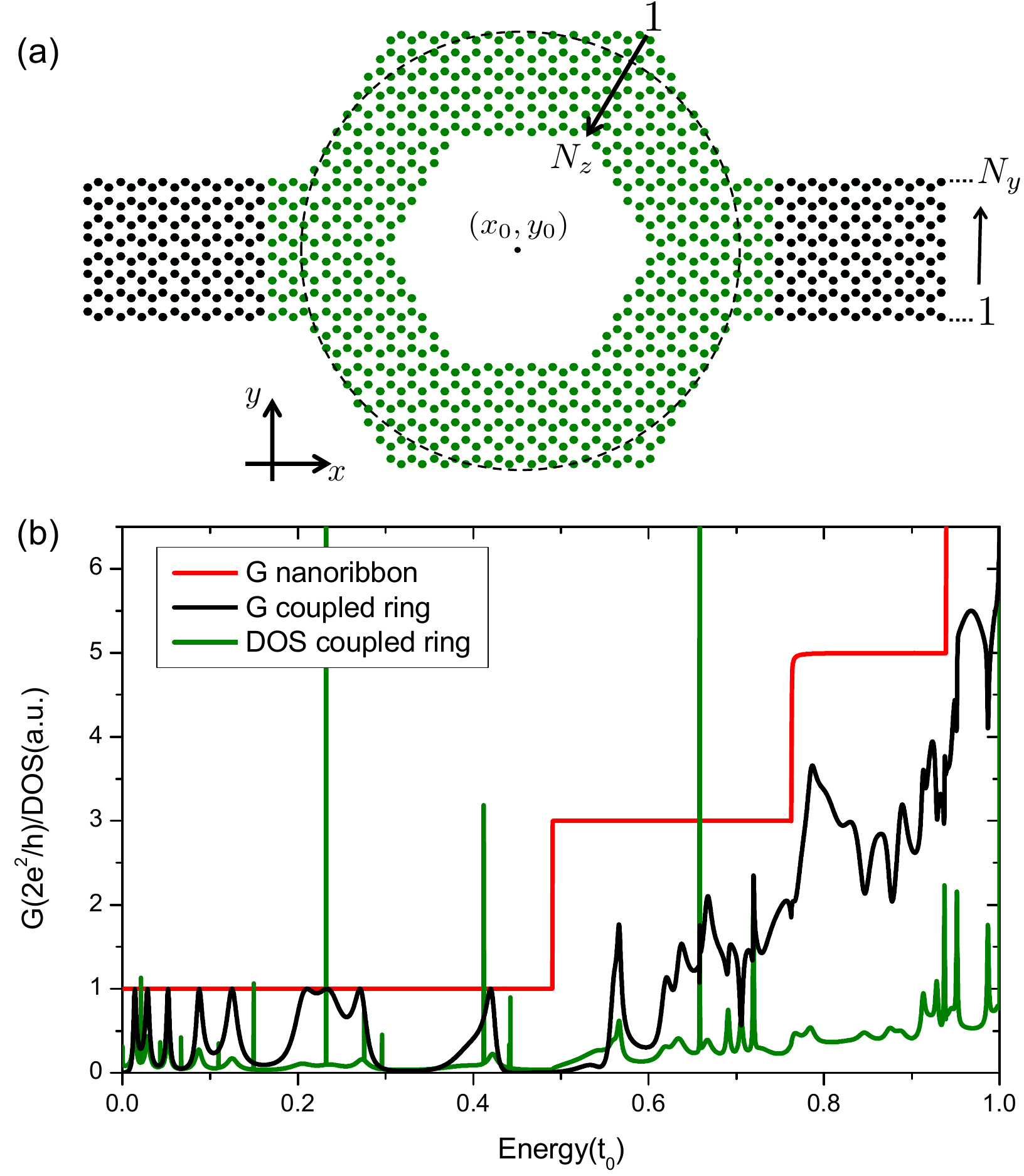}
\caption{(Color online) (a) Schematic view of the hexagonal ring connected to semi-infinite zigzag nanoribbons ($N_z=6$ and $N_y=8$). The coordinate system is shown in the lower left part and the mean external radius ($r\sim 17a$) of the ring is represented by the black dashed line. The structure shown contains 680 atoms in the central ring. 
(b) Conductance (black) and DOS (green) of the coupled system as a function of the Fermi energy for $\Phi=0$. The conductance for a perfect zigzag nanoribbon ($N_y=8$) is also plotted for comparison (red). Because of particle-hole symmetry, data is shown only for positive energies.} 
\label{fig1} 
\end{figure}

\section{Ring coupled to leads} 
A schematic representation of the model is shown in figure~\ref{fig1}(a). The geometry of the ring is important as long as it respects the underlying crystalline symmetry of the 2D graphene lattice and it was chosen to avoind change of edge orientations at the corners of the ring\cite{Adame}. The ring and leads are defined by the number of zigzag chains, $N_z$ and $N_y$, respectively. We use a standard,  single $\pi$-band nearest-neighbor tight-binding approximation to describe the central structure, with a Hamiltonian given by:
\begin{equation}
H_C=\sum_{<i,j>}t_{ij}c_i^{\dagger}c_j+\sum_{<i,j_{L(R)}>}t_{0}c_i^{\dagger}c_{j_{L(R)}} +h.c. \,\,\,,
\label{Hamiltoniano}
\end{equation}
where the fermion operator $c_i^{\dagger}(c_i)$ creates (annihilates) an electron in the i-th site and $t_0=-2.7 eV$ is the hopping parameter\cite{Saito}. The first term in equation~(\ref{Hamiltoniano}) represents the dynamics in the disconnected ring, with indices $i$ and $j$ running over all ring sites. The second term connects the ring to the leads, with $j_L$ and $j_R$ denoting sites on the left and right leads, respectively.  

An external magnetic field $\bf{B}=B \hat{z}$, permeating the entire ring, is included via the Peierls substitution. As a result the hopping parameter acquires a complex phase\cite{Saito},  as follows:  $t_{ij}=t_0e^{i\Delta \phi_{ij}}$, with $\Delta \phi_{ij}=2\pi (e/h)\int_{\bf{r}_{j}}^{\bf{r}_{i}}\bf{A}\cdot d\bf{r}$ and $\bf{r}_{i}$ and $\bf{r}_{j}$ being the nearest neighbors positions. We choose the Landau gauge $\bf{A}$=$(0,Bx,0)$, and measure the phase in units of the magnetic flux threading a single hexagonal plaquete, $\Phi/\Phi_0= 3a^2\sqrt{3}eB/2h$, with $a=1.42\AA$  being the interatomic distance. 

The DOS and Landauer conductance are calculated with the Green's function formalism \cite{Latge1,Datta}. Reservoirs are represented by the self-energy $\Sigma_{L(R)}$, obtained from each lead Green's function calculated with real-space renormalization techniques\cite{Carlos, Caio}. 

\begin{figure}[h!] 
\centering
\includegraphics[scale=0.28]{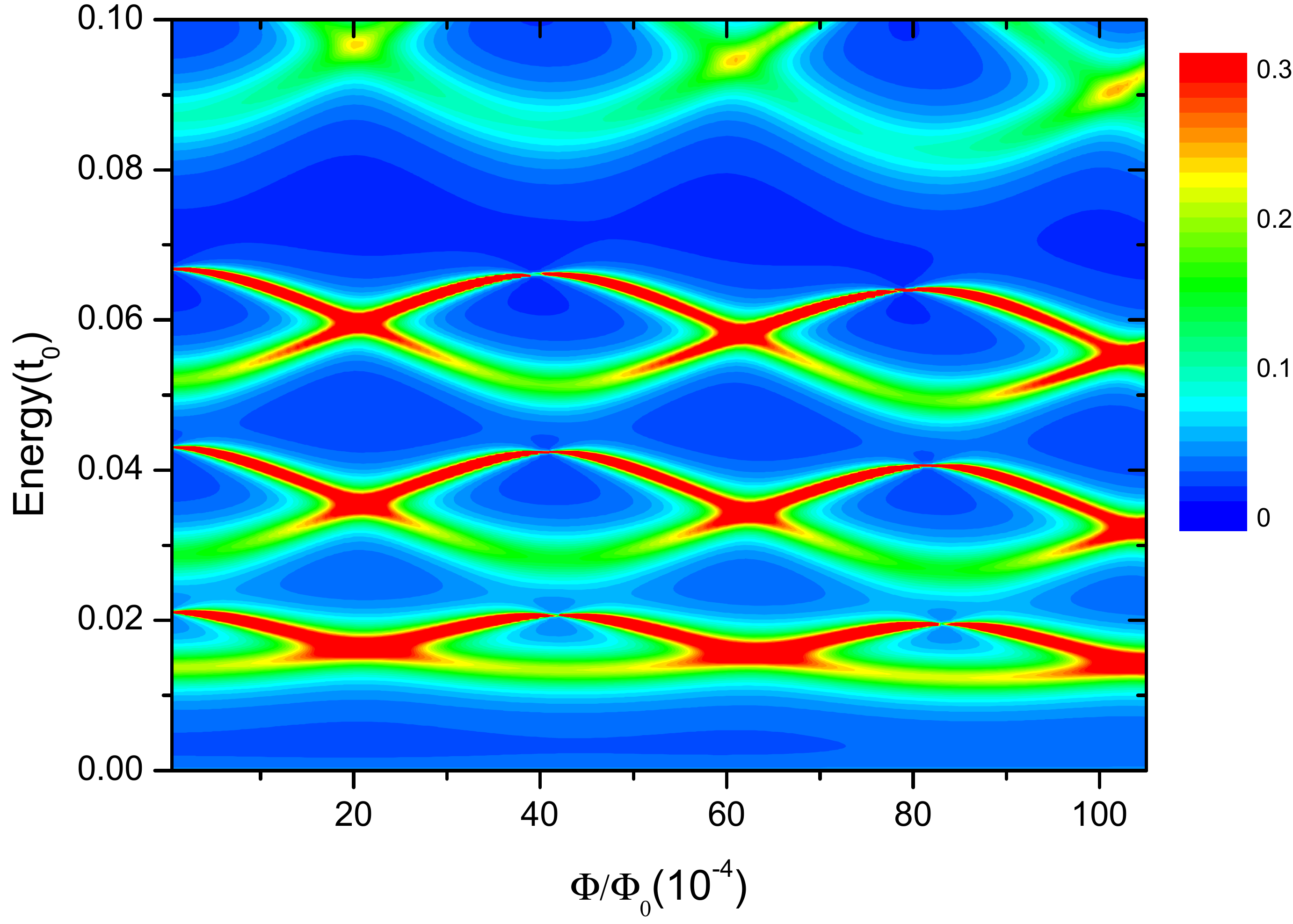}
\caption{(Color online) Contour plot of the total DOS as a function of  magnetic flux and Fermi energy. Color code represents DOS intensity.} 
\label{fig2} 
\end{figure}

Results for a typical system in the absence of magnetic field are displayed in figure~\ref{fig1}(b), where the energy dependence of the DOS (green line) and the conductance (black line) are shown. The conductance for a zigzag nanoribbon of the same width as the leads is also drawn for comparison (red line).  As expected, the low energy transport is caused by one-channel transmission. The figure shows some of the sharpest DOS peaks coinciding with minima in the conductance, thus not contributing to transport. This effect is also present at higher energies (see the second and third conductance plateaus for example).

\begin{figure}[h!]
\centering
\includegraphics[scale=0.57]{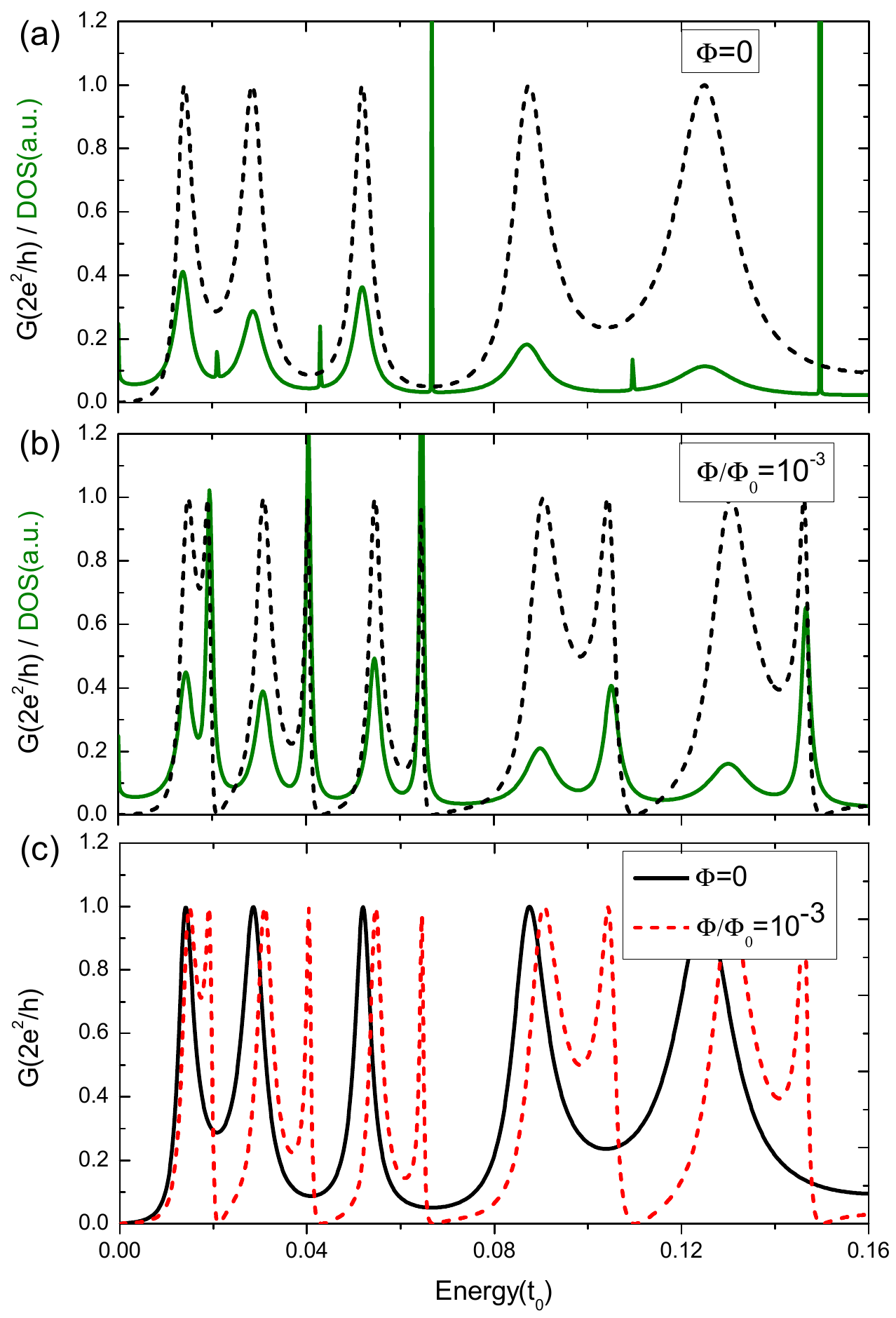}
\caption{(Color online) DOS (green solid line) and conductance (black dashed line) for: (a)  $\Phi=0$ and (b) $\Phi/\Phi_0 = 10^{-3}$. (c) Conductance comparison with and without external flux.} 
\label{fig3}
\end{figure}

We focus next in the energy range within the first conductance plateau to analyze the nature of these sharp peaks in detail. Results  for the  calculation of the total DOS as a function of the Fermi energy and the external magnetic flux are shown in figure~\ref{fig2}. The DOS is split into bands that exhibit an oscillatory dependence on the external flux, and evolve towards fully developed Landau levels in the high-field regime\cite{Morpurgo, Peeters, Xu}. Previous results\cite{Peter} have shown a similar dependence of the energy levels of the isolated ring, with this behavior being determined by the average radius. In contrast to the energy spectrum of an isolated hexagonal ring that exhibits six energy levels in each sub-band\cite{Peter}, the open structure contains sub-bands with two levels only. These levels display different broadening, indicating different coupling to the leads. Figure~\ref{fig2} shows that the reduced symmetry structure (from six-fold to two-fold) rendered by the connection to the leads, retains nevertheless those states that are compatible with both symmetries. Thus it follows that some of the states in the closed structure are not fundamentally altered by the symmetry reduction and persist in the open geometry.

\section{Fano resonances} 
Figure~\ref{fig3} shows results for conductance (dash black line) and DOS (full green line) with and without the external magnetic field. The open ring structure is insulating at zero energy as shown by the null conductance in the absence and presence of flux. Panel a) corresponds to zero field and show the features observed in figure~\ref{fig1} with greater detail. It is natural to associate the sharp peaks that coincide with minima in the conductance to localized states present in the open system. 

\begin{figure}[h!] 
\centering
\includegraphics[scale=0.55]{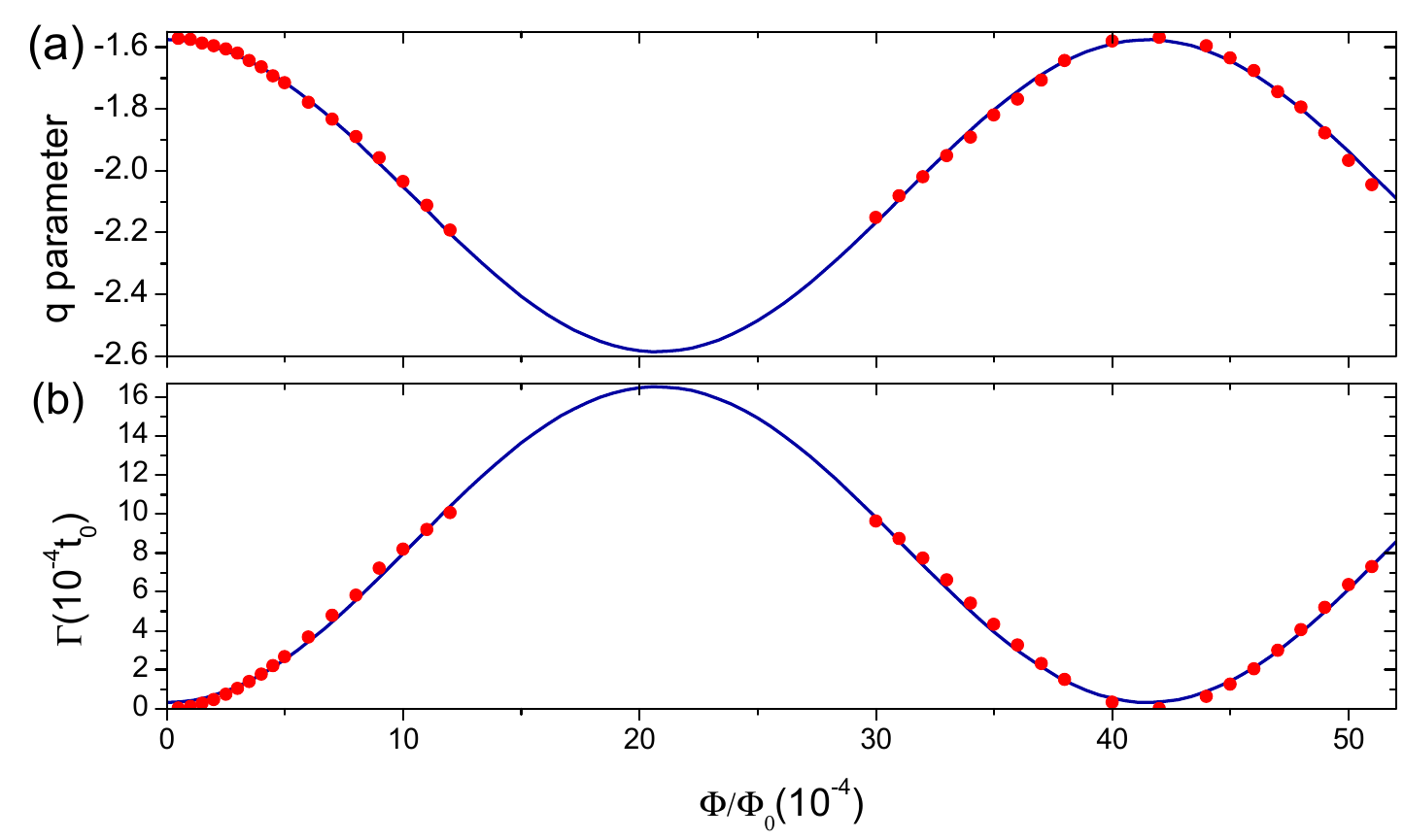}
\caption{(Color online)  (a) Fano parameter $q$ and (b) $\Gamma$ of function $\mathcal{G}(\epsilon)$ (equation~(\ref{Fano})) as a function of the magnetic flux. Dots (red) are fits for conductance values at the lowest energy resonance. Full line curve (blue) is a sinusoidal fitting of the data.} 
\label{fano}
\end{figure}

Figure~\ref{fig3}(b) shows results for a magnetic flux $\Phi/\Phi_0 = 10^{-3}$.  Note that the sharp peaks in the DOS of panel a) are broadened. The conductance develops an asymmetric peak-minima structure, with a zero value at the energies corresponding to the originally sharp peaks. This asymmetric line shape is the characteristic fingerprint of a Fano resonance\cite{Fano}. Figure~\ref{fig3}(c) shows the conductance with and without magnetic flux. In the presence of external flux there is a doubling of conductance peaks, a finding consistent with the existence of  a newly formed Fano resonance. A numerical fit of the conductance curve is obtained with the renormalized Fano expression: 
\begin{equation}
\mathcal{G}(\epsilon)=\frac{1}{1+q^2}\frac{(\epsilon+q)^2}{1+\epsilon} \,\,\,,
\label{Fano}
\end{equation}
being valid only if $|\epsilon|\ll1$, where $\epsilon=(E-E_0)/\Gamma$ is a reduced energy, $\Gamma$ is the resonance width, and $E_0$ is the resonance energy. The Fano fitting parameter $q$, a measure of the coupling between localized and extended states, acquires a periodic dependence on the external flux as shown in figure \ref{fano}. These results apply for all values of magnetic flux shown in figure~\ref{fig2}) except in the region where sub-band states mix. In these ranges a convolution of Fano and Breit-Wigner (symmetric-broadened peak) expressions is used to determine  the asymmetry degrees of the resonances\cite{Pedrox}.

\begin{figure}[h!] 
\centering
\includegraphics[scale=0.53]{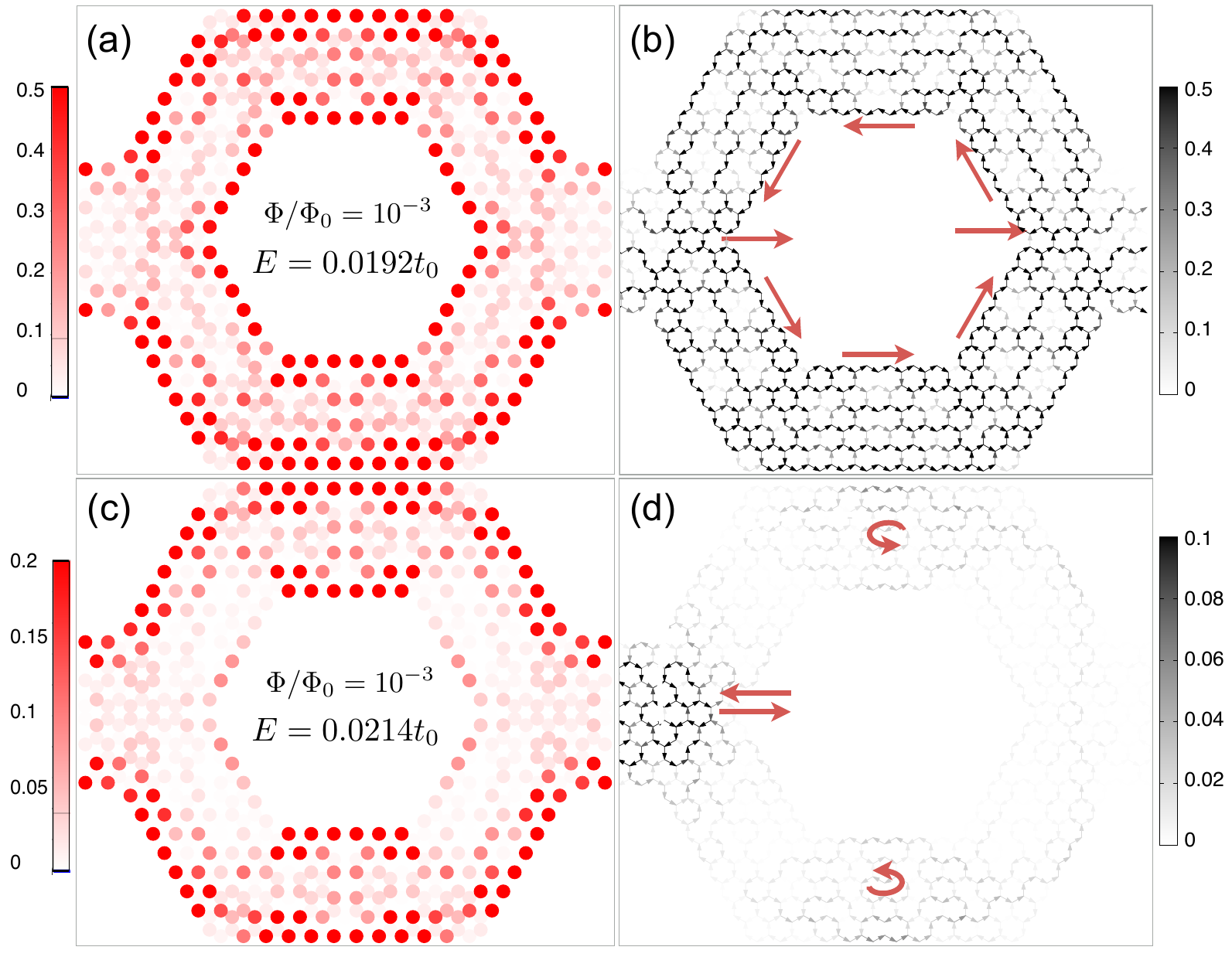}
\caption{(Color online) LDOS (left) and current density (right) mapping of the open ring for resonant $(E=0.0192t_0)$ and antiresonant states $(E=0.0214t_0)$. Here $\Phi/\Phi_0=10^{-3}$.  (a) and (b) correspond to the conductance peak, while (c) and (d) refer to the conductance minima. Notice different scales optimized for each case.}
\label{fig4} 
\end{figure}

To provide further confirmation that a Fano resonance is involved, we show results for current density patterns at energies corresponding to peaks and minima of the conductance \cite{Caio}. Our numerical calculations show that in the absence of magnetic field the current density splits equally over the two arms of the ring, for all energies studied. For a finite magnetic flux, in contrast, the current flows mostly through the upper or lower part of the ring, a consequence of the preferred circulation introduced by the field (not shown). Interestingly, the magnetic field generates two different current patterns for resonant $(E=0.0192t_0)$ and antiresonant states $(E=0.0214t_0)$ as shown in figure~\ref{fig4}.  Panel a) shows the LDOS for the resonant energy that extends over the whole ring, making possible perfect conductance. Panel b) shows the corresponding current  that effectively exists in both arms with a unique sense of circulation. In contrast, at the antiresonant energy (see figure~\ref{fig4}(c)), there is an enhanced LDOS at the central part of upper and lower arms of the ring,  consistent with the two-fold symmetry of the open structure. The current flow between the two terminals is completely suppressed as shown in figure~\ \ref{fig4}(d). Remarkably, there is a local charge circulation pattern at the central arms of the ring, that appears at a much smaller scale.   

The existence of finite LDOS at the value of the conductance minima confirms the existence of extended states inside the structure together with localized states at the upper and lower arms. These results provide additional evidence that the asymmetric conductance minima is produced by interference between extended and Fano resonant states.

\section{Strained ring}

 The model analyzed in previous sections describe suspended hexagonal graphene ring structures. However,  
structures available in current settings are commonly deposited on substrates that introduce additional effects. In particular, smoothly corrugated substrates cause local out of plane deformations that  produce strain in the system. To model it we consider a centro-symmetric Gaussian bump described by: 
\begin{equation}
h\left(r_{i}\right)=Ae^{-\left(r_{i}-r_{0}\right)^2/b^2},
\end{equation}
where $r_{i}$ represents an atomic site inside the ring  with coordinates $r_i=(x_i,y_i)$.  The Gaussian center $r_0=(x_0,y_0)$ coincides with the geometric ring center (figure~\ref{fig1}(a)). The deformation is included in equation~(\ref{Hamiltoniano}) as a modification in the hopping amplitude in the central structure\cite{Mehi}:
\begin{equation}
t^\prime_{ij}= t_{ij}e^{-\beta\left(l^\prime_{ij}/a-1\right)} \,,
\end{equation}
where the new atomic distances $l^\prime_{ij}$ are calculated using elasticity theory up to linear order on strain \cite{Ando,Juan,Ramon} and  $\beta=|\partial \log t_{0}/\partial \log a| \approx 3$. The new first-neighbor vectors are given by  $\vec{\delta}^\prime_{ij}=\vec{\delta}_{ij}.(I+\epsilon)$, with $I$ being the identity matrix and $\epsilon_{\gamma,\lambda}=\frac{1} {2}\partial_{\gamma}h\partial_{\lambda}h$ the strain tensor\cite{Landau}. We use the repeated greek index summation convention and $\gamma$ and $\lambda$ represent directions on the 2D plane. The strain parameter $\alpha=(A/b)^2$ is defined in terms of the amplitude $(A)$ and FWHM $(b)$ of the bump. Notice that strain fields introduce an effective pseudo-magnetic field that competes with the externally applied one.

\begin{figure}[h!] 
\centering
\includegraphics[scale=0.355]{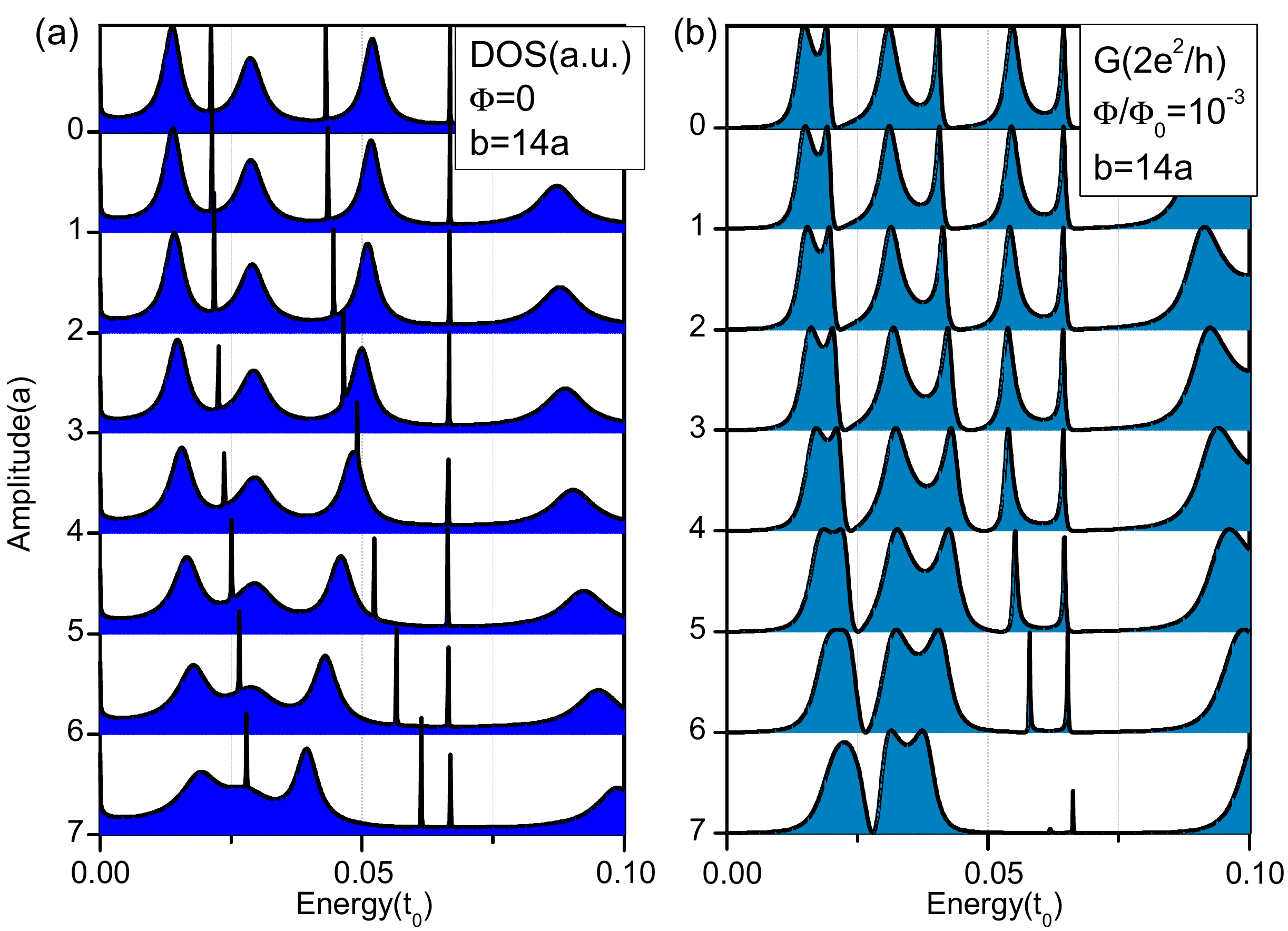}
\caption{(Color online) Strain effects on (a) DOS $(\Phi=0)$  and (b) conductance ($\Phi/\Phi_0=10^{-3}$), for different values of Gaussian deformation amplitude, and fixed FWHM $b =14a$. } 
\label{strain} 
\end{figure}

In figure~\ref{strain}(a) the DOS in the absence of external magnetic flux is shown for a strained graphene ring with varying Gaussian amplitude and fixed FWHM $(b=14a)$. The main effect of the deformation is to shift the position of various narrow peaks towards higher energy values. As an external flux is added, the strain promotes the Fano resonances to higher energies too, as shown in figure~\ref{strain}(b).  
These results highlight the persistence of localized states in the open ring structure and the robust effects of Fano resonances in the transmission in the presence of strain fields.  For higher strain values ($\alpha > 25\%$) other Fano resonances appear at low energies, even in the absence of an external magnetic flux. These are caused by new localized states produced by the strain fields at the location of maximum curvature of the deformation\cite{Ramon}. 

%

\section{Conclusions}
 We have studied a model for an hexagonal zigzag ring coupled to external contacts and showed that the structure contains localized states. These are remnants of the discrete states of the isolated ring that exist on the open structure due to the combination of the 2-fold and inversion symmetries of the graphene lattice\cite{chiral}. 
While these states do not contribute to the conductance, they can be detected by the application of an external magnetic flux that mixes them with the continuum background, generating Fano resonances. The Fano parameter $q$ characterizing the asymmetric interference peak in the conductance shows a periodic dependence on the applied flux. 

Contrarily to the Fano resonances previously addressed \cite{Adame}, that were provided by localized states induced by random edge disorder and spread exactly along the edge defects, the Fano resonance in our work are achieved in clean graphene rings, with no disorder. Interestingly, the kink formed at the corners of the hexagon acts as a defect-like, localizing states that are originally extended along the edges. The confinement of these states appears along the sides of the ring geometry and not at
the position of the kinks.

For ring structures deposited on corrugated substrates we analyzed the effect of strain on transport properties. We found that out-of plane deformations produce an overall shift in the energy of the resonances without affecting them otherwise. These results suggest that two terminal transport measurements in the presence of an external flux could be used to characterize small strain patterns in samples by analyzing shifts in conductance minima. In the limit of large strains, the dependence of the conductance minima with the curvature of the deformation suggests that transport could provide an alternative way to characterize strain profiles on samples.

\section{Acknowledgments}

We thank useful discussions with F. Adame, E. Bastos, C. Lewenkopf, L. Lima, F. Mireles and P. Orellana. This work was supported by CNPq, CAPES (2412110) and DAAD (D.F.); FAPERJ E-26/101.522/2010
(A.L.);  NSF No. DMR-1108285 (D.F., R.C-B. and N.S.), CONACYT, PAPIIT-DGAPA UNAM IN109911 (R.C-B).


\begin{thebibliography}{99}
 
\bibitem{Castro} A. H. Castro Neto, F. Guinea, N. M. R. Peres, K. S. Novoselov and A. K. Geim,  Rev. Mod. Phys. {\bf 81}, 109 (2009).

\bibitem{Trauzettel}  J. Schelter, P. Recher, B. Trauzettel, Solid State Communication {\bf 152}, 1411 (2012). 

 \bibitem{Morpurgo} P. Recher, B. Trauzettel, A. Rycerz, Ya. M. Blanter, C. W. J. Beenakker, and A. F. Morpurgo, Phys. Rev. B {\bf 76}, 235404 (2007).
  

\bibitem{Brey} T. Luo, A. P. Iyengar, H. A. Fertig, and L. Brey, Phys. Rev. B {\bf 80}, 165310 (2009).


\bibitem{Peter} D. A. Bahamon, A. L. C. Pereira, P. A. Schulz, Phys. Rev. B {\bf 79}, 125414 (2009).

\bibitem{Xu} M. M. Ma, J. W. Ding, and N. Xu, Nanoscale {\bf 1}, 387 (2009). 

\bibitem{Peeters} D. R. da Costa, A. Chaves, M. Zarenia, J. M. Pereira Jr, G. A. Farias, and F. M. Peeters, Phys. Rev. B {\bf 89}, 075418 (2014).

\bibitem{Daiara} D. Faria, A. Latg\'e, S. E. Ulloa, N. Sandler, Phys. Rev. B {\bf 87}, 241403(R) (2013).

 \bibitem{Wurm} J. Wurm, M. Wimmer, H. U. Baranger, and K. Richter, Semicond. Sci. Technol. {\bf 25}, 034003 (2010).

 \bibitem{Peeters1} Z. Wu, Z. Z. Zhang, K. Chang and F. M. Peeters, Nanotechnology {\bf 21}, 185201 (2010).

 \bibitem{Beenakker} A. Rycerz,  Acta Phys. Pol. A {\bf 115}, 322 (2009); A. Rycerz and C. W. J. Beenakker, arXiv:0709.3397. 

\bibitem{Russo} S. Russo, J. B. Oostinga, D. Wehenkel, H. B. Heersche, S. S. Sobhani,
L. M. K. Vandersypen, and A. F. Morpurgo, Phys. Rev. B {\bf 77}, 085413 (2008).

\bibitem{Ihn1} M. Huefner, F. Molitor, A. Jacobsen, A. Pioda, K. Ensslin, and T. Ihn, Phys. Status Solidi B {\bf 246}, 2756 (2009).

\bibitem{Ihn2}M. Huefner, F. Molitor, A. Jacobsen, A. Pioda, C. Stampfer, K. Ensslin, T. Ihn, New J. Phys. {\bf 12}, 043054 (2010).

\bibitem{Haug} D. Smirnov, H. Schmidt, and R. J. Haug, Appl. Phys. Lett. {\bf 100}, 203114 (2012).

\bibitem{Haug2} D. Smirnov, J. C. Rode, and R. J. Haug, arXiv:1407.8358.

\bibitem{Rahman} A. Rahman, J. W. Guikema, S. H. Lee, and N. Markovi\'c,  Phys. Rev. B {\bf 87}, 081401(R) (2013).

\bibitem{Hu} Y. Hu, M. Ruan, Z. Guo, R. Dong, J. Palmer, J. Hankinson, C. Berger and W. A. de Heer, J. Phys. D: Appl. Phys. {\bf 45}, 154010 (2012).

 \bibitem{Baringhaus} J. Baringhaus, M. Ruan, F. Edler, A. Tejeda, M. Sicot, A. T. Ibrahimi, Z. Jiang, E. Conrad, C. Berger, C. Tegenkamp, W. A. de Heer, Nature {\bf506}, 349 (2014).

\bibitem{Miro} A. E. Miroshnichenko, S. Flach, and Y. S. Kivshar, Rev. Mod. Phys. {\bf 82}, 2257 (2010).
 
 \bibitem{Boris} B. Luk'yanchuk, N. I. Zheludev, S. A. Maier, N. J. Halas, P. Nordlander, H. Giessen, and C. T. Chong, Nature Mater. {\bf 9}, 707 (2010).
 
\bibitem{FanoAharonov}  K. Kobayashi, H. Aikawa, S. Katsumoto, and Y. Iye,  Phys. Rev. Lett. 88, 256806 (2002).

\bibitem{Fano} U. Fano, Phys. Rev. {\bf124}, 1866 (1961).

\bibitem{Szafran} M. R. Poniedzialek and B. Szafran, J. Phys.: Condens. Matter, {\bf22} 465801 (2010).

\bibitem{Nowak} M. P. Nowak, B. Szafran, F. M. Peeters, Phys. Rev. B {\bf84}, 235319 (2011). 

\bibitem{Gonzales} J. W. Gonz\'ales, M. Pacheco, L. Rosales, and P. A. Orellana, Europhys. Lett. {\bf 91}, 6601 (2010).

\bibitem{Adame} J. Mun\'arriz, F. Dom\'inguez-Adame, and A. V. Malyshev, Nanotechnology, {\bf 22} 365201 (2011).

\bibitem{Adame1} J. Mun\'arriz, F. Dom\'inguez-Adame, P. A. Orellana and A. V. Malyshev, Nanotechnology, {\bf 23} 205202 (2012).

 \bibitem{Rosales} L. Rosales, M. Pacheco, Z. Barticevic, A. Latg\'e and P.  A. Orellana, Nanotechnology {\bf 19}, 065402 (2008).

\bibitem{Vozmediano} M. A. H. Vozmediano, M. I. Katsnelson, and F. Guinea,
Phys. Rep. {\bf 496}, 109 (2010).

\bibitem{Saito} R. Saito, M. S. Dresselhaus, and G. Dresselhaus,  {\it Physical Properties of Carbon Nanotubes}, (Imperial College Press, 1998).

\bibitem{Latge1} D. Grimm et al, Small 3, 1900 (2007);  C. Ritter,  R. B. Muniz, and A. Latg\'e, Appl. Phys. Lett. {\bf 104}, 143107 (2014). 
 
 \bibitem{Datta} S. Datta, {\it Electronic Transport in Mesoscopic System}, (Cambridge University Press, 1997).
 
 \bibitem{Carlos} C. Ritter, S. S. Makler, and A. Latg\'e, Phys. Rev. B {\bf 77}, 195443 (2008).

\bibitem{Caio} C. H. Lewenkopf and E. R. Mucciolo, Journal of Computational Electronics {\bf 12}, 203 (2013).

\bibitem {Pedrox} J. W. Gonz\'alez, M. Pacheco, L. Rosales , and P. A. Orellana, Phys. Rev. B {\bf 83}, 155450 (2011).

\bibitem{Mehi} D. A. Papaconstantopoulos, M. J. Mehl, S. C. Erwin, and
M. R. Pederson, {\it Tight-binding approach to Computational Material Science} (Materials Research Society, Pittsburgh,1998), p. 221.

\bibitem{Ando} H. Suzuura and T. Ando, Phys. Rev. B {\bf 65}, 235412 (2002).

\bibitem{Juan} F. de Juan, M. Sturla, and M. A. H. Vozmediano, Phys. Rev. Lett. {\bf 108}, 227205 (2012).

\bibitem{Ramon} R. Carrillo-Bastos, D. Faria, A. Latg\'e, F. Mireles, and N. Sandler, Phys. Rev. B {\bf 90}, 041411(R) (2014).

\bibitem {Landau} L. D. Landau and E. M. Lifshitz, {\it Theory of Elasticity}
(Pergamon Press, Oxford, 1970).

\bibitem{chiral} J. Mur-Petit and R. A. Molina, Phys. Rev. B {\bf 90}, 035434 (2014).





\end{thebibliography}
\end{document}